\documentclass{PoS}
\usepackage{colordvi}
\usepackage{graphicx}

 \leftline{\parbox{3cm}{\large\rm HU-EP-07/47 ADP-07-11/T651}
\vspace {1.0cm}}

\title{The Landau gauge gluon and ghost propagators in 4D $SU(3)$ gluodynamics
  in large lattice volumes}
\ShortTitle{Gluon and ghost propagators in large volumes}

\author{\speaker{I.~L.~Bogolubsky}$^{\hspace{1mm}a}$,
  E.-M.~Ilgenfritz$^b$, M.~M\"uller-Preussker$^{b}$, and A.~Sternbeck$^{c}$\\~\\
  $^{a}$ Joint Institute for Nuclear Research, 141980 Dubna, Russia\\
  $^{b}$ Humboldt Universit\"at zu Berlin, Institut f\"ur Physik, 12489 Berlin,
  Germany\\
  $^{c}$ CSSM, School of Chemistry \& Physics, University of Adelaide,
  SA 5005, Australia\\
  E-mail: \email{bogolubs@lxpub04.jinr.ru}, \email{ilgenfri@physik.hu-berlin.de},
  \email{mmp@physik.hu-berlin.de},
  \email{andre.sternbeck@adelaide.edu.au}
}

\abstract{We present recent results on the Landau gauge gluon and
  ghost propagators in $SU(3)$ pure gauge theory at Wilson $\beta=5.7$
  for lattice sizes up to $80^4$ corresponding to physical volumes up
  to $(13.2 {\rm~fm})^4$. In particular, we focus on finite-volume and
  Gribov-copy effects. We employ a gauge-fixing
  method that combines a simulated annealing algorithm with finalizing
  overrelaxation. We find the gluon propagator for the largest
  volumes to become flat at $q^2 \sim 0.01 {\rm~GeV}^2$. Although
  not excluded by our data, there is still no clear indication of a
  gluon propagator tending towards zero in the zero-momentum
  limit. New data for the ghost propagator are reported, too. }

\FullConference{The XXV International Symposium on Lattice Field Theory \\
July 30 -- August 4, 2007 \\ Regensburg, Germany}

\begin{document}

\vspace{-0.3cm}
\section{Introduction}
\vspace{-0.1cm}

Presently, there is an intensive exchange of results and opinions
between groups carrying out analytical and numerical studies of
infrared QCD. This ongoing research focuses in particular on the
infrared behavior of the Landau-gauge gluon and ghost propagators.
The latter is intimately related to the
confinement~\cite{Zwanziger,KuOj,AvS}. What makes
analytical predictions possible also in the non-perturbative
sector of the theory is the possibility to write down
a hierarchy of Dyson-Schwinger equations (DSE) connecting
propagators and vertices. Under rather mild assumptions the
hierarchy can be truncated. However, not all assumptions have been
thoroughly checked. Note that recently the full system of Landau
gauge DSE has been solved without any truncations within the
asymptotic infrared region with a power ansatz for
all Green functions involved \cite{AFLP}. Numerically, the
propagators can be studied from first principles in terms of Monte
Carlo (MC) simulations of lattice QCD. It is worth to compare the
lattice results with the asymptotic power-like behavior, and with
numerical DSE solutions found in finite volumes \cite{FMPS07}.
It is interesting to see whether there remain differences as the
infinite-volume limit is approached.

On the lattice we approximate the gluon propagator as the MC average
\begin{equation}
  D^{ab}_{\mu\nu}(q) = \langle \tilde{A}^{a}_{\mu}(\hat{q})
  \tilde{A}^{b}_{\nu}(-\hat{q}) \rangle =
  \delta^{ab} \left(\delta_{\mu\nu} -\frac{q_{\mu}q_{\nu}}{q^2}\right)
  \frac{Z_{gl}(q^2)}{q^2}
\end{equation}
with the gluon field
$A_{x+\hat{\mu}/2,\mu}
= (1/2iag_0) ( U_{x,\mu} - U^{\dagger}_{x,\mu} )_{\rm traceless}$
transformed into Fourier space. The lattice momenta
$\hat{k}_{\mu}= 2~\pi~k_\mu/L_{\mu}$ with integer
$k_{\mu} \in (-L_{\mu}/2, +L_{\mu}/2]$ are related to their physical values
by $q_{\mu}=(2/a) \sin(\pi~k_{\mu}/L_{\mu})$.

The ghost propagator in momentum space at non-zero $q^2$ is defined by
double Fourier transformation
\begin{equation}
  G^{ab}(q) = \sum_{x,y} \left\langle
    {\rm e}^{-i \hat{k} \cdot (x-y)}~[M^{-1}]^{ab}_{x,y} \right\rangle
  = \delta^{ab}\frac{Z_{gh}(q^2)}{q^2} \;.
\end{equation}
Practically, this is done by an inversion of the Faddeev-Popov (F-P)
matrix $M^{ab}_{x,y}$ using a conjugate-gradient algorithm with plane
waves as sources. The Faddeev-Popov operator in terms of the Landau
gauge-fixed links is
\begin{equation}
M^{ab}_{xy} = \sum_{\mu} \mathfrak{Re}~\mathrm{Tr}~\Big[ \{T^a,T^b\}(U_{x,\mu} +
U_{x-\hat{\mu},\mu})\delta_{xy} -2T^b T^a U_{x,\mu}\delta_{x+\hat{\mu},y} -2T^a
T^b U_{x-\hat{\mu},\mu}\delta_{x-\hat{\mu},y} \Big] \; ,
\end{equation}
with $T^a = \lambda^a/2$ ($\lambda^a$ are the Gell-Mann matrices).
The functions $Z_{gl}(q^2)$ and $Z_{gh}(q^2)$ are called dressing functions of
the respective propagator.

In Landau gauge the gluon and ghost dressing functions are predicted to
follow the simple power laws \cite{vonSmekal:1997is}
\begin{equation}
Z_{gh}(q^2) \propto (q^2)^{-\kappa} \quad\mbox{~and}\quad
       Z_{gl}(q^2) \propto (q^2)^{2\kappa} \; .
\end{equation}
in the asymptotic regime $q^2 \to 0$. Thereby,
both exponents are related to some $\kappa$ which,
under the assumption that the ghost-gluon
vertex is infrared-regular, takes a value of about
$\kappa=0.596$ \cite{LS02Zwan}. That is, the gluon propagator is
predicted to decrease towards lower momenta and to vanish at $q^2=0$.
At which scale this asymptotic behavior sets in cannot be concluded
from those studies, however.

A truncated system of DSE formulated on a 4D torus and numerically
solved \cite{FMPS07} predicts a specific finite-volume behavior
which, at the first glance, looks quite similar to earlier lattice
results obtained in particular by some of us \cite{SIMPS}.
Characteristic deviations for momenta $q \sim 1/L$ of the gluon
and the ghost propagator from the momentum dependence of the
respective propagators at infinite volume should be expected. In
order to check the DSE predictions we decided to evaluate the
gluon and the ghost propagator for increasingly large symmetric
lattices. We have measured the gluon and ghost propagators for
configurations generated with the Wilson gauge action at fixed
$\beta=5.7$ on $56^4$, $64^4$, $72^4$ and $80^4$ lattices. Note
that the latter corresponds to a volume of about $(13.2
{\rm~fm})^4$. Comparing results from either analytic or
numerical approaches for varying 4-volume will hopefully allow (i)
to conclude for which momenta data on both propagators are
reliable, and (ii) to estimate the order of magnitude of distortion
of the momentum dependence due to finite-size and Gribov-copy
effects. This paper presents first results of this study.
\vspace{-0.3cm}

\section{Gauge fixing}
\vspace{-0.1cm}

To fix the Landau gauge, we apply to all links a gauge transformation $g \in G$
($G = SU(3)$) mapping
$U_{x,\mu} \to {}^gU_{x,\mu}=g_x U_{x,\mu} g^{\dagger}_{x+\hat{\mu}}$,
with the aim to maximize a gauge functional
\begin{equation}
F_U[g] = \frac{1}{12V} \sum_{x,\mu}
\mathfrak{Re}~\mathrm{Tr}~{}^gU_{x,\mu} \; ,
\label{fu}
\end{equation}
or, more exactly, to find the global maximum of $F_U[g]$ ~\cite{Zwanziger}.
In this work $g_x \in G$ is considered as a periodic field on all lattice
sites. To find the global maximum in practice is a complicated problem
which becomes exceedingly time consuming with increasing lattice volume.
Starting from an initial random gauge transformation $g_x$ one
generally arrives at one of many local maxima of $F_U[g]$. The
corresponding gauge-fixed configurations are
called Gribov copies. They all satisfy the differential gauge condition
$\partial_{\mu} A_{\mu} = 0$ together with the additional necessary
condition that the Faddeev-Popov operator has a positive spectrum (apart
from its 8 trivial zero modes). With increasing volume, the copies
become dense with respect to the value of the functional (\ref{fu})
and the spectral density of the F-P operator near zero grows
\cite{Sternbeck:2005vs}.

One way to suppress the effect of the Gribov ambiguity is to find
$N_{\rm copy}$ local maxima of $F_U[g]$ and to choose among them
the ``best'' one (``bc''), which possesses the largest value of
$F_U[g]$. The underlying idea is that the maximal value of the
local maxima approaches the global maximum of $F_U[g]$, and the
distortion of gauge-dependent observables, computed on such
copies, vanishes in the limit $N_{\rm copy}\to \infty$. Such
studies normally use the overrelaxation (OR) technique to
search for the maximum of $F_U[g]$. They have been carried out in
\cite{SIMPS,BIMMP,BBMMP,IMMPSSB} and have shown the Gribov-copy
effect to become weaker with growing lattice extension $L$ -- in
accordance with Zwanziger's conjecture~\cite{Zwanziger}. This
suggests that it is tolerable to restrict gauge-fixing
computations on large lattices $(L \ge 48)$ to one gauge copy
only\footnote{This is called ``first copy'' (``fc'') in a
multi-copy approach.}. Our simulations of the $SU(3)$ propagators
at $L=56$ were carried out using an OR algorithm with the
overrelaxation parameter set to $\alpha=1.70$. The number of
gauge-fixing
(GF) iterations did not exceed $10^4$ in most of the cases.
However, we find a considerable slowing down of the OR GF
process on a $64^4$ lattice. This was one of the reasons why we
switched from using OR to a simulated annealing (SA) algorithm.
SA, also known as a ``stochastic optimization method'' has been
proven to be highly effective in coming close to the global
maximum in various problems of different nature with multiple
local maxima. It was proposed in \cite{KGV,Cerny} and has found
numerous applications in various fields of science. The idea of
applying SA for gauge fixing has been first put forward and
realized in the case of maximally Abelian gauge in \cite{BBMPP}.
The method is designed to keep the system long enough pending in a
region of simultaneous attraction by many local maxima during a
quasi-equilibrium process undergone by the ``spin system'' formed by
$g_x$ interacting through the fixed $\{U_{x,\mu}\}$ field with
$F_U[g]$ as energy. The temperature $T$ of the spin system is
decreased by small $T$-steps between updates in a range of $T$
where the penetrability of functional barriers strongly changes.
Theoretically, when infinitely-slow cooling down to $T=0$, the SA
algorithm finds the global maximum with 100\% probability. For
complicated systems with large numbers of degrees of freedom and
of functional local extrema, e.g., for GF on large lattices, we
have to restrict the number of $N_{\rm iter}$ $T$-steps of cooling
the system from $T_{\rm max}$ to $T_{\rm min}$ to, say, $O(10^4)$.
Within these limits we can still try to attain an as high value of
the functional studied (in our case, $F_U[g]$) as possible. Note
that GF with SA requires a finalizing OR in order to satisfy
transversality $\partial_{\mu} A_{\mu} = 0$ with a given high
precision.

For the $SU(3)$ case we chose $T_{\rm max}$ such that it
leads to a sufficiently large mobility in the functional space.
The final temperature $T_{\rm min}$ was taken low enough that
the subsequent OR was not slowed down while penetrating further
functional barriers. This is witnessed by the check that the
violation of the differential gauge condition, $(\partial_{\mu}
A_{\mu})^2$, monotonously decreases until the machine precision is
reached (stopping criterion) in almost all cases. In practice, for
$L=56,64,72$ and $80$ we restricted ourselves to one copy,
and carried out from $5 \times 10^3$ to $15 \times 10^3$ heatbath
(HB) sweeps of SA with 4 microcanonical sweeps after each HB one.
We checked that the smaller $T$-steps are done in between,
the higher the local maxima being reached. Finally, we note that a
linear decrease in $T$ seems not to be the optimal choice.
$T$-schedules with smaller $T$-steps close to $T_{\rm max}$ and
larger $T$-steps at the end (with $N_{\rm iter}$ fixed) lead to
higher $F_U[g]$-values (after completing the full SA procedure).
\vspace{-0.3cm}

\section{Ghost propagator results}
\vspace{-0.1cm}

\begin{figure}
\begin{center}
\includegraphics[height=7.6cm,width=7.8cm,angle=270]%
{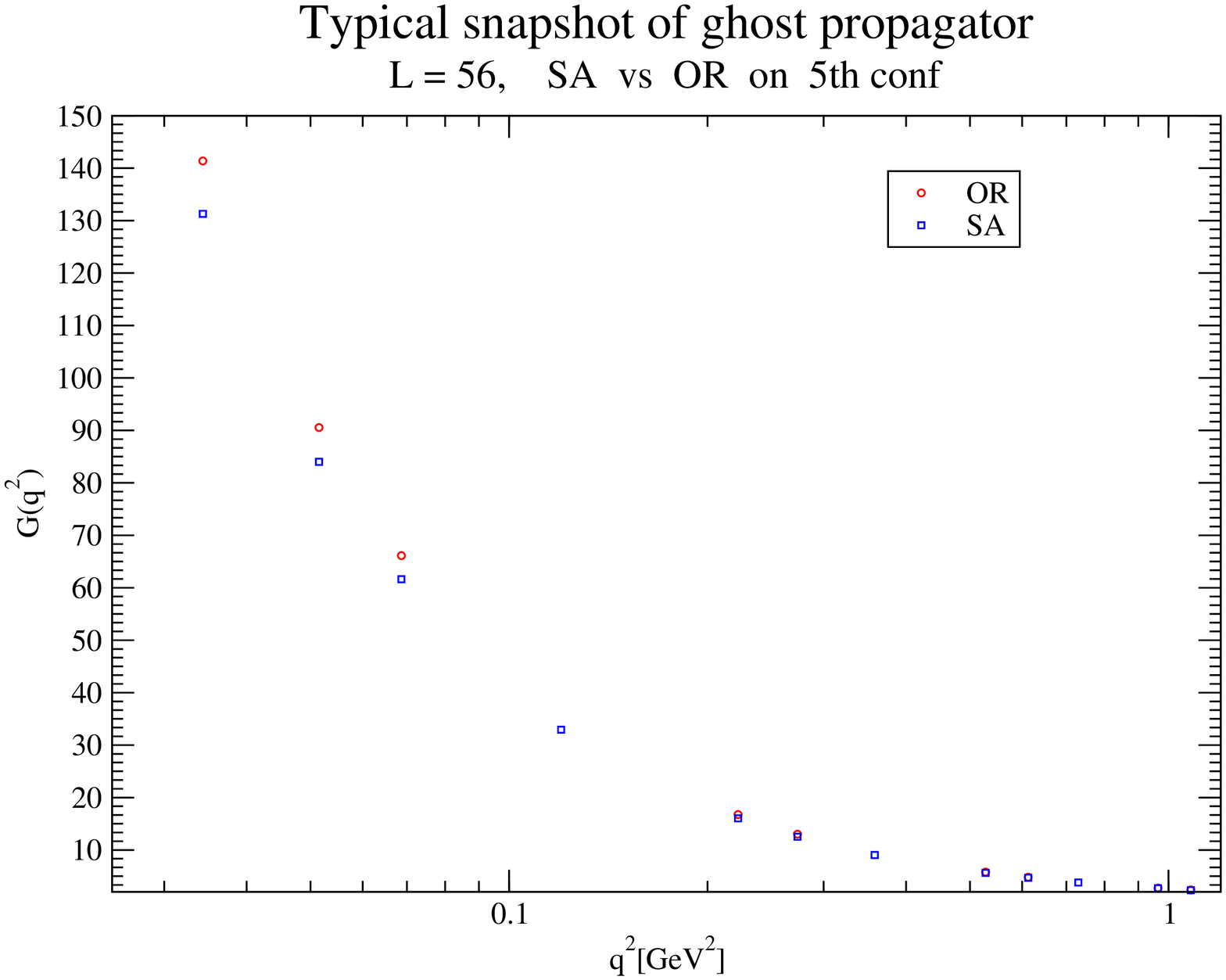} \hspace{-0.3cm}
\includegraphics[height=7.6cm,width=7.8cm,angle=270]%
{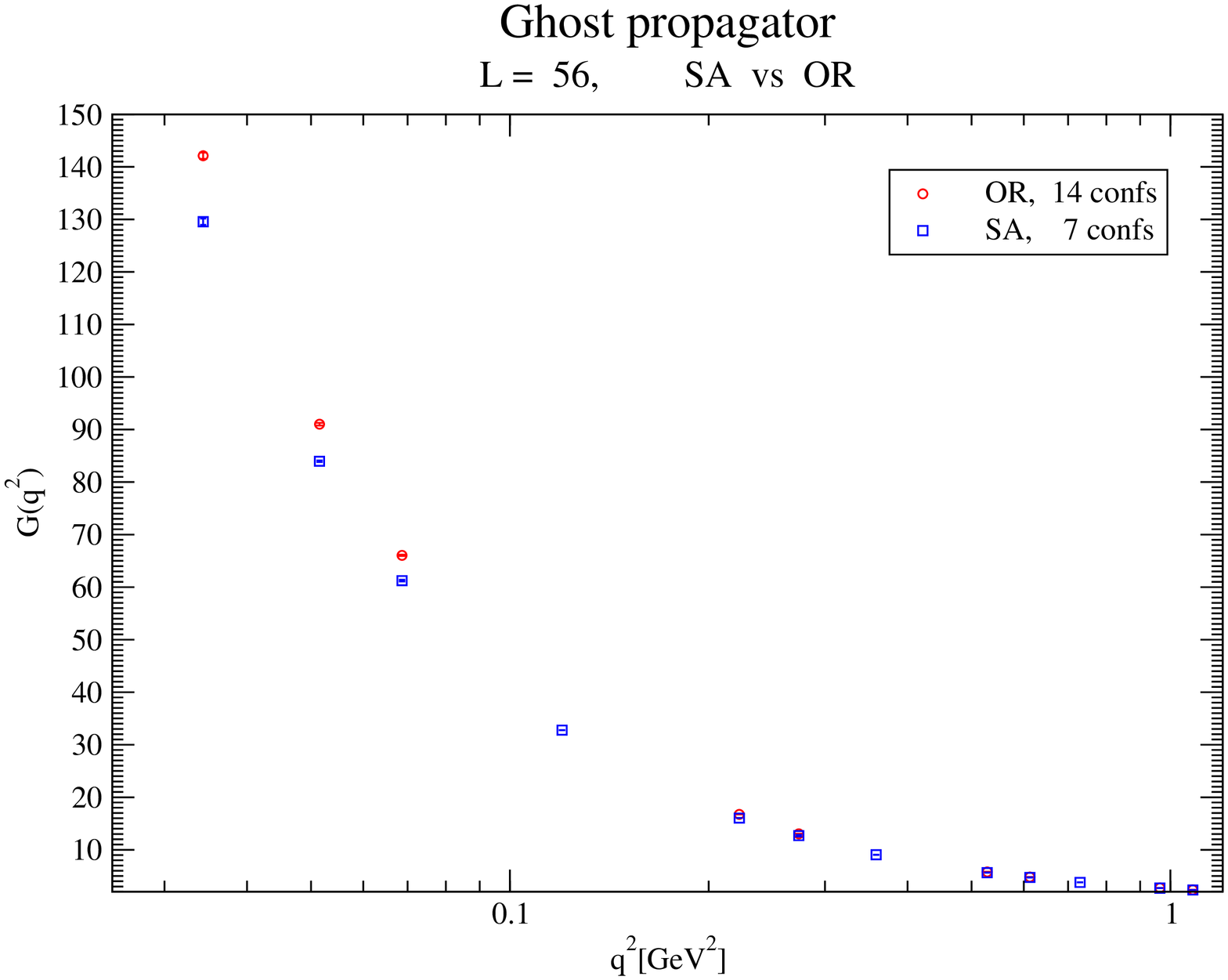}
\end{center}
\vspace{-1.0cm}
\caption{Ghost propagator results: the influence of the gauge
fixing algorithm on the propagator, calculated for a typical
single gauge field configuration (left), on the averaged
propagator (right).} \label{fig:fig1}
\end{figure}

The $SU(3)$ ghost propagator at $\beta=5.7$ is shown as a function
of $q^2$ in Fig.~\ref{fig:fig1}; on the left hand side for a
single $56^4$ configuration, simply comparing results after either OR
or SA gauge fixing, and on the right hand side as an average over
14 configurations in the case of OR, and over 7 configurations
using SA gauge fixing. The influence of the gauge-fixing method,
here through the emerging copy, can be seen only for the three
lowest momenta at this lattice size. In general, one notices that
the higher the gauge functional, the lower the estimates of the ghost
propagator at the smallest momenta. This comparison
(Fig.~\ref{fig:fig1}) demonstrates that the problem of Gribov
copies does still exist for $L=56$, resulting in maximally $10\%$
difference of ghost propagators at lowest momenta. Note that this
result cannot be directly compared to our previous studies
\cite{BIMMP,SIMPS,BBMMP,IMMPSSB} of Gribov-copies effect, in which
the ``fc-bc'' comparison was used to assess the Gribov
ambiguity. A detailed check of Zwanziger's conjecture on the
weakening of the Gribov problem with an increase of the lattice
volume (using the SA vs.\ OR ``one-copy'' comparison) requires
further studies both for smaller and larger lattices.

\begin{figure}
\centering
\includegraphics[height=8cm,angle=270]{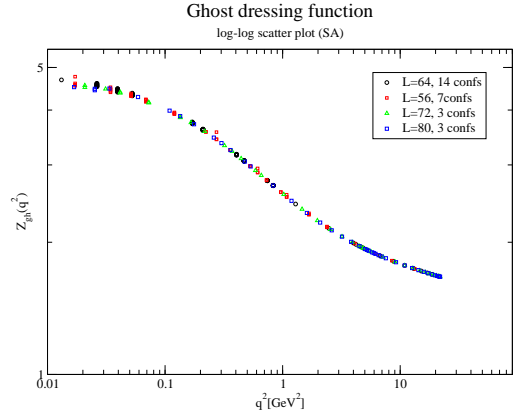}
\caption{Scatter plot for the ghost dressing function from
different lattice sizes and configurations, generated at
$\beta=5.7$ and gauge-fixed with SA.}
\label{fig:fig2}
\end{figure}

In Fig.~\ref{fig:fig2} a scatter plot is shown of the ghost
dressing function for a broad range of momentum, combining data
obtained for 7 configurations on a $56^4$ lattice, 14
configurations on a $64^4$ lattice, 3 configurations on a $72^4$
and 3 configurations on a $80^4$ lattice, all thermalized at
$\beta=5.7$. The gauge field configurations were produced with a
heat-bath algorithm applying O(1000) thermalization sweeps in
between. We consider only momenta surviving a cylinder cut with
$\Delta{\hat q} =1 $ \cite{LSWP99}. For all lattice sizes GF has
been carried out with the SA algorithm. Surprisingly, all the
values for the ghost propagator fall perfectly on one universal
curve (within $1 \%$ accuracy), besides those for the 2 smallest
momenta.
 The results, especially those found at $L=80$, show that a true IR
exponent $\kappa$ cannot yet be defined or does not exist at all.
This is at variance with the asymptotic DSE prediction
$\kappa=0.595$ \cite{FMPS07} and also with $\kappa=0.2$ motivated
by thermodynamic considerations in \cite{ChZ07}. The latter
estimate of $\kappa$ is based on the required cancellation of
gluon and ghost contributions to the pressure, that otherwise were
building up a Stefan-Boltzmann law, in the confinement phase. Note
that downward deviations of the data at lowest momentum for each
physical box size $V$ from the infinite-volume curve of $G(q^2)$
are predicted by the DSE approach on a finite torus~\cite{FMPS07}.
However, we do not find such deviations.

\vspace{-0.3cm}

\section{Gluon propagator results}
\vspace{-0.1cm}

\begin{figure}
\centering
\includegraphics[height=8.0cm,angle=270]{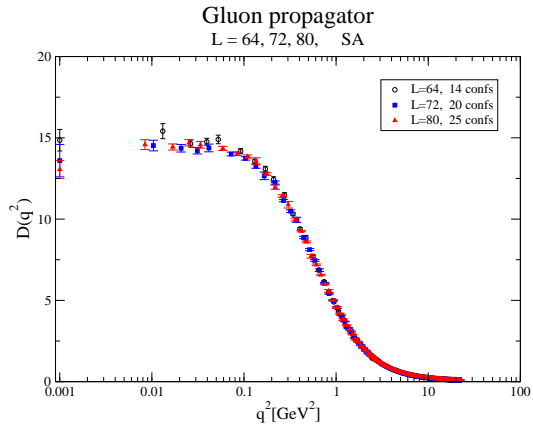}
\caption{The gluon propagator from different lattice sizes at
  $\beta=5.7$. The data points drawn at $q^2 = 0.001$ represent the
  zero-momentum gluon propagator $D(0)$.} \label{fig:fig3}
\end{figure}

Fig.~\ref{fig:fig3} shows data for the gluon propagator computed
for three different lattice sizes at $\beta=5.7$. There, the gauge
was fixed with the SA algorithm. At the present stage, the data
favor a non-vanishing gluon propagator at zero momentum, as there
is no sight of a different behavior even at the largest lattice
volume available to us. Also, the decrease of the zero-momentum
propagator $D(0)$ upon increasing the volume seems to become less
with bigger $V$. The $64^4$ data for $D(q^2)$ resemble the pattern
of overshooting deviations from an universal function of momentum
known from the DSE solutions on a finite torus~\cite{FMPS07},
though. If the lowest two or three momenta were removed from the
plot, the picture would be less convincing in favor of an emerging
plateau. Better statistics and data on even larger symmetric
lattices (with $L > 80$) will make us more confident in this.

\vspace{-0.3cm}

\section{Discussion}
\vspace{-0.1cm}

We conclude that using the SA technique for the purpose of gauge
fixing considerably facilitates simulations of ghost and gluon
propagators in Landau gauge on large lattices. We find that
SA-based computations of the ghost propagator seem to be less
affected by statistical fluctuations compared to other
calculations where OR is used. Additionally, estimates of the ghost
propagator at low momenta are systematically lower than those obtained
after simple OR.

A continuous decrease in slope in the ghost dressing function below
0.4~GeV does not conform to a simple power-law ansatz. Therefore, any
attempts to extract infrared exponents from lattice data seem to be
premature at the present stage. Qualitatively, the same behavior is
seen for the case of $SU(2)$ (see \cite{SVLW}) and also in the DSE
solutions on a torus \cite{FMPS07}. However, the effects of finite
volumes are much less than expected from there though. The same we
find for the gluon propagator which we cannot confirm to be
infrared-decreasing even at volumes larger than $(13\textrm{fm})^4$.

Future analytical (DSE and renormalization group) studies and
lattice simulations, optionally including also $\mathbb{Z}(3)$
flip operations into the GF procedure \cite{BBMMP}, at even larger
volumes will hopefully help to resolve or explain the existing
discrepancies between the lattice and analytical findings.

 \vspace{-0.3cm}

\section*{Acknowledgements}
\vspace{-0.1cm}

Simulations were done on the MVS-15000BM and MVS-50000BM at the Joint
Supercomputer Centre (JSCC) in Moscow and on the IBM pSeries 690 at HLRN.
This work was supported by joint grants DFG 436 RUS 113/866/0 and RFBR
06-02-04014. Part of this work is supported by DFG under contract FOR 465
Mu 932/2-4, and by the Australian Research Council.

\end{document}